\documentstyle[epsfig,graphics]{mn}

\begin{document}

\newcommand{\Zsolar}{\mbox{\,$\rm Z_{\odot}$}}
\newcommand{\Ysolar}{\mbox{\,$\rm Y_{\odot}$}}
\newcommand{\Msolar}{\mbox{\,$\rm M_{\odot}$}}
\newcommand{\etal}{{et al.}\ }
\newcommand{\ang}{\mbox{$\rm \AA$}}
\newcommand{\xs}{$\chi^{2}$}
\newcommand{\ls}{{\tiny \( \stackrel{<}{\sim}\)}}
\newcommand{\gs}{{\tiny \( \stackrel{>}{\sim}\)}}
\newcommand{\teff}{T$_{eff}$}

\title[F stars, metallicity, and ages of red galaxies]{F stars, metallicity, and the ages of
red galaxies at {\bf $ z > 1$}}
\author[L.A. Nolan \etal.]
{L.A. Nolan$^{1}$, J.S. Dunlop$^{2}$, R. Jimenez$^{3}$, $\&$ A.F. Heavens$^{2}$
\\
$^{1}$School of Physics and Astronomy, University of Birmingham, Edgbaston, Birmingham, B15~2TT\\
$^{2}$Institute for Astronomy, University of Edinburgh, Blackford Hill, Edinburgh, EH9~3HJ\\
$^{3}$Physics and Astronomy Department, Rutgers University, 136 Frelinghuysen Road, Piscataway, NJ
08854-8019, USA.}

\date{Submitted for publication in MNRAS}

\maketitle
  
\begin{abstract}

We explore whether the rest-frame near-ultraviolet spectral region, 
observable in high-redshift galaxies via optical spectroscopy, 
contains sufficient information to allow the degeneracy between 
age and metallicity to be lifted. We do this by first testing the ability 
of evolutionary synthesis models to reclaim the correct metallicity 
when fitted to the near-ultraviolet spectra of F stars of known 
(sub-solar and super-solar) metallicity. F stars are of particular interest
because the rest-frame near-ultraviolet spectra of the oldest known
elliptical galaxies at $z > 1$ appear to be dominated by F stars near
to the main-sequence turnoff.

We find that, in the case of the F stars, where the HST ultraviolet
spectra have high signal:noise, model-fitting with metallicity allowed
to vary as a free parameter is rather successful at deriving the
correct metallicity. As a result, the estimated turnoff ages of these
stars yielded by the model fitting are well constrained.  Encouraged
by this we have fitted these same variable-metallicity models to the
deep, optical spectra of the $z \simeq 1.5$ mJy radio galaxies 53W091
and 53W069 obtained with the Keck telescope.  While the age and
metallicity are not so easily constrained for these galaxies, we find
that even when metallicity is allowed as a free parameter, the best
estimates of their ages are still $\geq$ 3 Gyr, with ages younger than
2 Gyr now strongly excluded. Furthermore, we find that a search of the
entire parameter space of metallicity and star formation history using
MOPED leads to the same conclusion. Our results therefore continue to
argue strongly against an Einstein-de Sitter universe, and favour a
$\Lambda$-dominated universe in which star formation in at least these
particular elliptical galaxies was completed somewhere in the redshift
range $z = 3 - 5$.

\end{abstract}


\section{Introduction}

Deep optical spectroscopy with $8-10$-metre class telescopes can
overcome many of the problems associated with estimating the ages of
high-redshift galaxies from cruder broad-band data. In particular, as
demonstrated by Dunlop et al. (1996), Spinrad et al. (1997) and Dunlop
(1999), high-quality spectral data not only provide reassurance that
one is observing starlight, but allow reddening-independent age
estimation based on the strength of specific spectral features. These
observational advances have greatly enhanced the prospects for
reliable age-dating of high-redshift stellar
populations. Consequently, recent years have seen renewed interest in
clarifying and attempting to resolve the remaining sources of
uncertainty. As is often the case when an area of high-redshift
astronomy finally enters the regime of quantitative astrophysics, the
dominant sources of uncertainty transpire to be problems that have
dogged work at lower redshift for many years. In this case the main
outstanding issues are disagreements over the details of stellar
evolution, and the well-known degeneracy between derived age and
metallicity (Worthey 1994).

Of course the potential severity of both, and especially the former of
these effects is expected to be a function of age and observed
spectral region. In particular, as described by Magris \& Bruzual
(1993), and reaffirmed in the study of the red, $z \simeq 1.5$ galaxy
LBDS 53W091 (Dunlop et al. 1996; Spinrad et al. 1997), it is in the
rest-frame near-ultraviolet (near-UV), at ages less than 5 Gyr, that
disagreements between the predictions of alternative models of
spectral evolution should be minimised.  This is because for such ages
it is expected that the near-UV light should be dominated by the
output of stars near the turnoff point of the `well-understood' main
sequence (e.g. Nolan et al. 2001). However, the ensuing controversy
over the precise age of 53W091 (Bruzual \& Magris 1998; Yi et
al. 2000), and the resulting confusion over the wider implications for
cosmology in general, have highlighted the need for a closer
inspection of the uncertainties which continue to afflict galaxy
age-estimation.
 
In an attempt to clarify the key issues we have now begun to study
both the extent to which various evolutionary synthesis models differ
in their evolutionary predictions, and the extent to which rest-frame
near-UV data may allow the age-metallicity degeneracy to be lifted.
In Nolan et al. (2001), we checked the time evolution of the
solar-metallicity (\Zsolar) population synthesis models of several
authors, by comparing the ability of the main-sequence components of
these models to reproduce the near-UV solar spectrum of the Sun at its
anticipated main-sequence turn-off age.

In this paper we explore whether, given data of sufficient quality,
variable-metallicity stellar population models can extract the correct metallicity of a population simply from fitting to the near-UV spectral region
accessible through optical spectroscopy of high-redshift galaxies. 

We test whether the UV spectral region contains sufficient information
to allow reliable recovery of the metallicity in two stages. Firstly,
the single stellar atmosphere model spectra of Jimenez et al. (1998)
are fitted to the HST ultraviolet spectra of two well-studied,
non-solar metallicity, main-sequence stars, at a resolution $\lambda /
\Delta \lambda = 30,000$ (Heap et al., 1998a), and a typical
signal-to-noise per pixel of 50, before re-binning (Heap et al.,
1998b). These are kindly supplied by Sally Heap (Heap, private
communication). Secondly, variable- and mixed-metallicity versions of
the stellar population models of Jimenez et al.(1998)
are fitted to the same two stellar spectra. These stars are HR 4683,
an F4V star, with sub-solar metallicity, and HR 4688, an F8V star with
super-solar metallicity (Edvardsson et al., 1993). The motivation for
using the spectra of these particular stars is that the rest-frame
near-UV spectra of the oldest known galaxies at $z \simeq 1.5$
(e.g. 53W091 and 53W069) mimic the spectra of main-sequence F stars
(F5V - F9V; Dunlop 1999).

This is, of course, a rather one-sided test, but it is still
potentially very instructive. Specifically, if the single stellar
atmosphere models cannot extract the correct metallicity from the
high-quality near-UV data available for these F-stars, then it is
probably safe to assume that lifting the age-metallicity degeneracy in
high-redshift galaxies is currently impossible. Conversely, if the
single atmosphere models can yield the correct metallicity for these
F-stars, then that provides some hope that the age-metallicity
degeneracy can be lifted in high-redshift galaxies, given near-UV data
of sufficient quality. In addition, if the variable- and
mixed-metallicity stellar population model results are consistent with
the results from fitting the single stellar atmospheres, then we may
be confident that, at young ages, main-sequence stars do indeed
dominate the UV flux of stellar populations, that the time-evolution
of the non-solar metallicity population models is correct, and that we
can constrain metallicity even with the greater flexibility allowed in
the mixed-metallicity population models. We may then be confident that
ages and metallicities recovered for the high-redshift galaxies are
reliable.

However, such optimism must be tempered by the knowledge that the
spectrum of a high-redshift galaxy is unlikely to be as clean as that
of the test-case F-stars, given the obvious likelihood of a mixture of
metallicities and ages in a substantial stellar population.
Nevertheless, as we report in this paper, the results of fitting the
variable/mixed metallicity models of Jimenez et al. (1998) to the
near-UV spectra of the two F-stars are sufficiently encouraging to
motivate the re-fitting of the Keck spectra of LBDS 53W091 and 53W069
with metallicity allowed to vary as a free parameter.
   
The layout of the paper is as follows. In section 2 we describe the
models we have used, and the procedures by which they have been fitted
to the data. In section 3 we present the results of fitting single,
variable metallicity, model stellar atmosphere spectra, single (but
variable) metallicity stellar population models, and mixed-metallicity
stellar population models (comprising a free mix of seven different
metallicities) to the HST near-UV spectra of the F stars. Then, in
section 4, we present the results of fitting the same stellar
population models to the optical (rest-frame near-UV) spectra of the
$z \simeq 1.5$ galaxies LBDS 53W091 and 53W069, and discuss the extent
to which such modelling allows the age-metallicity degeneracy to be
lifted. Finally, as a check on whether our assumption of single burst
models is appropriate, we have also fitted the spectra of these two
galaxies using MOPED (Heavens et al., 2000), which allows the star
formation history to be unconstrained. Our conclusions, including
revised constraints on the ages of these red galaxies, are summarised
in section 5.
 
\section{Models and model fitting}

\subsection{The models}

The stellar population evolutionary synthesis models utilised in this
paper are the instantaneous starburst models of Jimenez et al. (1998),
and the model single stellar atmospheres are those used in creating
these population models.  The reliability and consistency of the solar
metallicity (\Zsolar $ = 0.02$) version of these population models
were confirmed by Nolan et al. (2001).  We now also possess non-solar
metallicity models, for both the stellar population models and the
single stellar atmosphere models, with metallicities 0.01\Zsolar,
0.2\Zsolar, 0.5\Zsolar, 1.5\Zsolar, 2.5\Zsolar and 5.0\Zsolar.

The stellar interior library of the Jimenez et al. (1998) models is
computed using the latest version of the JMSTAR15 code (Jimenez et
al. 1996, Jimenez \& McDonald 1996). This uses the OPAL95 opacities
(Iglesias \& Rogers 1996) for stellar temperatures \gs\ 6000K, and
Alexander's opacities for temperatures $<$ 6000K. The mixing-length
theory of Baker \& Temensvary (1966), modified to allow for the limit
of optically thin convective elements (Mihalas 1978), is used for the
treatment of convective energy transport. The model grid is
calculated for a range of metallicities, Z $=$ 0.0002 $-$ 0.1, with
solar metallicity, \Zsolar\ $=$ 0.02, and the solar helium fraction,
\Ysolar\ $=$ 0.28. The mixing-length parameter, $\alpha$, is
calibrated by a comparison of the computed solar model with the
observed effective temperature, \teff, age, luminosity and radius of
the Sun. In addition, a consistency check was carried out, using the
position of the red giant branch (RGB) in globular clusters with known
metallicity (Jimenez et al. 1996).

Stellar masses range from 0.1 to 120 \Msolar, with step size, $\Delta$
M $=$ 0.05 \Msolar, from 0.1 to 3 \Msolar, and $\Delta$ M $=$ 1 \Msolar\
thereafter. The models used in this work have a Salpeter IMF (Salpeter
1955). The tracks are evolved all the way from the contracting Hayashi
track to the planetary nebula (PN) phase, except for the most massive
stars, where the evolution is followed until carbon ignition. All
tracks are computed using solar-scaled abundances.

Stellar evolutionary tracks are computed using the boundary conditions
set by the theoretical photospheric models calculated by Jimenez et
al. (1998). Therefore, for each calculated point on the track, the
associated stellar spectrum is known, and there is no need to
introduce the uncertainties which arise in calibrating real stellar
spectra to theoretical isochrones. Semi-convection arises as a natural
consequence of this modelling method.

JMSTAR15 computes the whole evolution of a star in a single run,
including the helium core flash, through the thermally-pulsing
asymptotic giant branch (TPAGB), ejecting a PN at the end. This is
possible due to the adaptive mesh of grid points, which allows very
rapid stages of evolution to be modelled in detail. The SSP code
includes a semi-empirical algorithm for computing the evolution of the
RGB, horizontal branch (HB) and asymptotic giant branch (AGB). This is
more realistic than a purely numerical code, and faster, but allows
the same degree of accuracy. The AGB was modelled as in J$\o$rgensen
(1991).

A semi-empirical algorithm is also used in the calculation of
mass-loss. The fast computation of post-main sequence (post-MS) stages
allowed several values of the mass-loss efficiency parameter, ($\eta$
in Reimers' formula, Reimers 1975), and mixing-length parameter,
$\alpha$, to be investigated. A realistic mean value of $\eta$ is
then found by matching the observed HB mass distribution (Jimenez et
al. 1995).

The stellar spectra are those of the Kurucz (1992) library of
theoretical stellar atmosphere models for stars with \teff\ $>$ 6000
K. For cooler stars, theoretical photospheres are computed by Jimenez
et al. (1998), using the improved (Helling et al. 1996) MARCS code
(Gustafsson et al. 1975), which includes molecular opacities.

\subsection{The fitting processes}

Firstly, we have explored the results of simply fitting each of the
seven single stellar atmosphere models in turn to the F star spectra,
at all possible ages, to identify the best-fitting single value of
metallicity, and the resulting inferred age. Secondly, the same
process is carried out for the seven single-metallicity stellar
population models, for both the F star spectra and the high-redshift
galaxy spectra. Finally, we have combined the seven single-metallicity
stellar population models to produce a composite model (of single age)
in which mean metallicity is allowed to vary by varying the fractional
contributions made by each of the seven different metallicity
components. We have fitted the mixed-metallicity model to the spectra
of the F stars, as well as the high redshift galaxies' spectra, to
confirm that fitting this model allows us to reclaim the correct
metallicity.

\subsection{Fitting the single stellar atmosphere and single-metallicity population models}
The best fit was determined for each single-metallicity model
(single-atmosphere and stellar population), i.e 0.01\Zsolar,
0.2\Zsolar, 0.5\Zsolar, \Zsolar, 1.5\Zsolar, 2.5\Zsolar and
5.0\Zsolar, in the following way. First, the data were re-binned to
the same spectral resolution as the models, with the new flux
calculated as the mean flux per unit wavelength in the new bin. Both
the re-binned and model fluxes were then normalised to a mean flux per
unit wavelength of unity across the wavelength range $3000 - 3500
\ang$ for the galaxy fits, and across the range $3000 - 3117 \ang$ for
the F-star fits (because of the shorter wavelength range of the F-star
spectra). The model age and normalisation were then varied as free
parameters until \xs\ was minimised. The wavelength range $2445-2465 \
\ang$ was excluded from the fits to all objects. When included, this
feature contributed to ~1/3 of the total \xs\ for both the F stars. As
this was felt to unfairly bias the fitting process for these objects,
the area was masked from the fit for all the objects, and may indicate
a failure of the models in this range.

In order to produce the contour plots for the single-metallicity
stellar atmosphere and stellar population model fits in sections 3 and
4, the results were refined by carrying out two one-dimensional cubic
spline interpolations, one for age, and one for metallicity, between
the calculated points on the age-metallicity grid.

\subsection{Fitting mixed-metallicity models}

A mixed-metallicity model was constructed from the 0.01\Zsolar,
0.2\Zsolar, 0.5\Zsolar, \Zsolar, 1.5\Zsolar\, 2.5\Zsolar\ and
5.0\Zsolar\ stellar population models of Jimenez et al. (1998). The
relative contributions of the different metallicity models were
allowed as free parameters. In this section, we assume that all the
component populations were formed at the same time, so that the one
other free parameter in the fit was the age of the complete
population. In section 4.3, we lift this constraint and explore what
happens when the stellar populations are formed at arbitrary times.


\begin{figure}
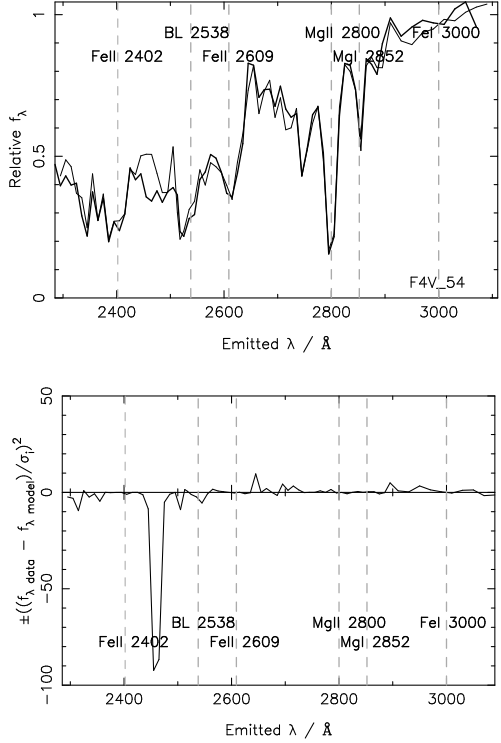

\centerline{\epsfig{file=F4V_541Zatmos.epsf.004,width=4.7cm,angle=-90,clip=}}
\vspace*{0.4cm}
\centerline{\epsfig{file=F4V_54xspec.1Zatmos.epsf.0.004,width=4.7cm,angle=-90,clip=}}

\caption{Top: the best-fitting, single stellar atmosphere model (thin
line, 0.2 \Zsolar\ model at 4 Gyr) to the spectrum of the F4V star, HR
4683 (thick line). Bottom: \xs\ residuals. Where $ ( {\rm f}_{\lambda data} < {\rm f}_{\lambda model} )$, the quantity plotted is $- ( ({\rm f}_{\lambda data} - {\rm f}_{\lambda model}) / \sigma )^{2}$. The central wavelengths of
the mid-UV spectral features identified by Fanelli et al. (1992),
lying within this wavelength range, are marked on the spectra with
dashed lines. See \S3.1 for discussion, and table 2 for values of
$\chi^2_{\nu}$.}

\end{figure}

\begin{figure}
\centerline{\epsfig{file=F8V+21.1Zatmos.epsf.0.05,width=4.7cm,angle=-90,clip=}}
\vspace*{0.4cm}
\centerline{\epsfig{file=F8V+21xspec.1Zatmos.epsf.0.05,width=4.7cm,angle=-90,clip=}}

\caption{Top: the best-fitting, single stellar atmosphere model (thin
  line, 2.5 \Zsolar\ model at 1.5 Gyr) to the spectrum of the F8V
  star, HR 4688 (thick line). Bottom: \xs\ residuals. Where $ ( {\rm
    f}_{\lambda data} < {\rm f}_{\lambda model} )$, the quantity
  plotted is $- ( ({\rm f}_{\lambda data} - {\rm f}_{\lambda model}) /
  \sigma )^{2}$. The central wavelengths of the mid-UV spectral
  features identified by Fanelli et al. (1992), lying within this
  wavelength range, are marked on the spectra with dashed lines. See
  \S3.1 for discussion, and table 1 for values of $\chi^2_{\nu}$.}

\end{figure}


\begin{figure}
\centerline{\epsfig{file=F4V_54.epsf.1,width=4.7cm,angle=-90,clip=}}
\vspace*{0.4cm}
\centerline{\epsfig{file=1zcontf4v.epsf,width=4.7cm,angle=-90,clip=}}

\caption{Top: the best-fitting, single-metallicity stellar population
model (thin line, 0.2 \Zsolar\ model at 3 Gyr) to the spectrum of the
F4V star, HR 4683 (thick line). Bottom: contours of constant reduced
$\chi^2$ ($\chi^2_{\nu}$) as a function of metallicity and age after
interpolation. The contour levels contain 68\%, 90\% and 95.4\%
relative likelihood. The shaded regions represent the estimates of
Edvardsson et al., with estimated errors $\pm$ 30\% in age and $\pm$
0.10 dex in metallicity. \xs\ residuals are shown in figure 5. See
\S3.2 for discussion, and table 2 for values of $\chi^2_{\nu}$.}

\end{figure}

\begin{figure}
\centerline{\epsfig{file=F8V+21.epsf.4,width=4.7cm,angle=-90,clip=}}
\vspace*{0.4cm}
\centerline{\epsfig{file=1zcontf8v.epsf,width=4.7cm,angle=-90,clip=}}

\caption{Top: the best-fitting, single-metallicity stellar population
model (thin line, 1.5 \Zsolar\ model at 3 Gyr) to the spectrum of the
F8V star, HR 4688 (thick line). Bottom: contours of constant
$\chi^2_{\nu}$ as a function of metallicity and age after
interpolation. The contour levels contain 68\%, 90\% and 95.4\%
relative likelihood. The shaded regions represent the estimates of
Edvardsson et al., with estimated errors $\pm$ 30\% in age and $\pm$
0.10 dex in metallicity. \xs\ residuals are shown in figure 6. See
\S3.2 for discussion, and table 2 for values of $\chi^2_{\nu}$.}

\end{figure}

The mixed-metallicity model flux was built from the unnormalised
single-metallicity model fluxes, and then normalised to a mean flux
per unit wavelength of unity in the same regions as for the
single-metallicity model fits, so that

 \[ F_{7Z,\lambda,t} = const \sum_{i=1}^{7}  X_{Z_{i}} f_{Z_{i},\lambda,t} \] 
where $F_{7Z,\lambda,t}$ is the mixed, seven-metallicity model flux
per unit wavelength in the bin centered on wavelength $\lambda$ at
age, $t$ Gyr; $f_{Z_{i},\lambda,t}$ is the flux per unit wavelength
$\lambda$ at age, $t$ Gyr, for the single metallicity model with $Z =
Z_{i}$, where $i = 1,7$, and $X_{Z_{i}}$ is the fractional
contribution (by baryonic mass) of the model with $Z = Z_{i}$ to the
total population.

The best-fitting age, normalisation and the value of each $X_{Z_{i}}$
were determined by \xs\ minimisation, with the wavelength range
$2445-2465 \ \ang$ masked from the fit, as it was in the single
stellar atmosphere and single stellar population fits. The entire
parameter space of the seven-dimensional age-metallicity hyper-cube
was searched, so that the best fits quoted in sections 3 and 4 are
those parameter values which correspond to the point on the
age-metallicity grid with the minimum \xs.

\subsection{Errors}
The errors on the binned spectral data points for the LBDS galaxies
have been derived from the propagation of the original observational
errors in the Keck optical spectroscopy (Dey et al., 2002, in
preparation). Using these errors, statistically acceptable fits were
achieved. For the F-star spectra, the error adopted for each re-binned
data point was simply the standard error in the mean for that bin. For
HR 4683 (the low-metallicity star), a formally acceptable fit was
obtained using these errors. However, for HR 4688 (the
high-metallicity star), this did not prove possible. There are a
number of reasons why this might occur (e.g. inadequacy of the
super-solar models $ - $ e.g. Kotoneva, Flynn \& Jimenez (1998).

\subsection{Marginalising the mixed-metallicity results}

The contours of constant relative likelihood discussed in \S 3.3 and
4.2 are plotted with respect to mean metallicity and age. As it is
possible for more than one combination of single-metallicity models to
result in the same mean metallicity, the metallicity distribution was
marginalised before plotting.

\section{Comparison with F-star spectra}

\subsection{Single stellar atmospheres}

\begin{table}

\begin{center}
\begin{tabular}{rrrr}

\\

\hline

 \\
	{object}  & { Z/{\mbox{\,$\rm Z_{\odot}$}}} & {Best fit age/Gyr} & { \xs$_{\nu}$ }  	\\

 \\

\hline  

\\

F4V  & 0.01 & 8   & 2.47        \\
     & 0.2  & 4   & 1.59        \\
     & 0.5  & 3   & 1.73        \\
     & 1.0  & 2   & 2.43 	\\
     & 1.5  & 1.5 & 2.71 	\\
     & 2.5  & 1   & 4.00 	\\
     & 5.0  & 0.5 & 8.31      	\\

\\

\hline  

\\

F8V  & 0.01 & not on MS  &      \\
     & 0.2  & 5   & 3.19        \\
     & 0.5  & 3   & 3.21 	\\
     & 1.0  & 2.5 & 2.97 	\\
     & 1.5  & 2   & 2.58 	\\
     & 2.5  & 1.5 & 2.40        \\
     & 5.0  & 1.5 & 3.20        \\

\\

\hline  

\end{tabular}
\end{center}

\caption{The results of fitting the near-ultraviolet spectra of the F-stars using the suite of seven single-metallicity single-atmosphere stellar models. The best fits for each of the single-metallicity stellar atmosphere models corresponding to the minimum $\chi^2_{\nu}$ are shown. }
\end{table}

In table 1 and figures 1 and 2, the results of fitting the seven
single-atmosphere models, with metallicities 0.01\Zsolar, 0.2\Zsolar,
0.5\Zsolar, \Zsolar, 1.5\Zsolar, 2.5\Zsolar and 5.0\Zsolar\ to the HST
spectra of the two F stars are presented. The best-fit single-star
model to the F4V star spectrum has a metallicity of 0.2\Zsolar, and an
age of 4 Gyr. For the F8V star, the best-fit single-star model has an
age of 1.5 Gyr, with a metallicity 2.5\Zsolar. 

Edvardsson et al. (1993) estimated the relative iron abundances
([Fe/H] = Z/\Zsolar, where $\Zsolar = 0.02$) of these two stars by
fitting model stellar atmospheres to spectral lines observed at
optical wavelengths.  They estimated the uncertainty in the derived
metallicities to be, at most, 0.10 dex. Based on these optical lines,
they found the F4V star to have a metallicity of 0.29\Zsolar, and the
F8V star to have a metallicity of 1.62\Zsolar. 

Edvardsson and coworkers also derived ages for these stars from fits
in the $T_{eff} - {\rm log }\ g $ plane to VandenBerg (1985)
isochrones. They found an age of 4.4 Gyr for the F4 dwarf, HR 4683,
and an age of 2.1 Gyr for the F8 dwarf, HR 4688. However, they expect
the uncertainties in the relative ages of the stars in their sample to
be $\sim25$\%, with the possibility that the absolute errors are
larger. 

Our results from fitting the single stellar atmosphere models are in
excellent agreement with these, entirely independently obtained, age
and metallicity estimates. Encouraged that the UV spectral region contains sufficient data to determine both age and metallicity, we proceed to fit the stellar population models.

\subsection{Single-metallicity stellar population models}

Attempting to fit complete stellar population models to the spectra of
single stars might seem like a strange thing to do. However, having
established that the individual input stellar spectra seem to work
well in the near-UV regime, extending the model fitting to complete
stellar populations offers a means to test the extent to which the
near-UV light from a complete stellar population is indeed dominated
by the dwarfs near the MS turnoff. It is also important to test
whether, with the addition of stars spanning a wide range in mass and
evolutionary stage, the theoretical UV spectrum can still identify the
correct metallicity of our adopted turn-off population (as represented
by the single F-star spectrum) given data of this quality. Finally,
expanding the parameter space still further to include a mix of
stellar populations of different metallicity, it is of interest to
test whether these stellar spectra still contain enough information to
allow the age and metallicity to both be well constrained.  

In figures 3 and 4, we show the results of fitting each of the seven
single-metallicity stellar population models in turn to the HST
spectra of the F stars HR 4683 and HR 4688 respectively. The first
plot in each figure shows the best-fitting model spectrum superimposed
on the observed near-UV stellar spectrum. The second plot shows
contours of constant reduced $\chi^2$ ($\chi^2_{\nu}$) as a function
of metallicity and age. The contour levels contain 68\%, 90\% and
95.4\% relative likelihood. The results are presented in table 2.

\begin{table}

\begin{center}
\begin{tabular}{rrrr}

\\

\hline

 \\
	{object}  & { Z/{\mbox{\,$\rm Z_{\odot}$}}} & {Best fit age/Gyr} & { \xs$_{\nu}$ }  	\\

 \\

\hline  

\\

F4V  & 0.01 & 13 & 10.5        \\
     & 0.2  & 3  & 1.68         \\
     & 0.5  & 3  & 1.72         \\
     & 1.0  & 1  & 2.79 	\\
     & 1.5  & 1  & 4.48 	\\
     & 2.5  & 0.7& 5.29 	\\
     & 5.0  & 0.3& 8.45 	\\

\\

     
\hline  

\\

  
F8V  & 0.01 & 13 & 21.7         \\
     & 0.2  & 13 & 2.86         \\
     & 0.5  & 13 & 2.55 	\\
     & 1.0  & 4  & 2.52 	\\
     & 1.5  & 3  & 2.35 	\\
     & 2.5  & 1  & 2.47         \\
     & 5.0  & 0.7& 3.76 	\\

\\

\hline  

\\

LBDS 53W069  	& 0.01 & 13 &  2.69     \\
	 	& 0.2  & 13 &  0.92     \\
     		& 0.5  & 9  &  0.92    	\\
     		& 1.0  & 4  &  1.04 	\\
     		& 1.5  & 3  &  1.02 	\\
     		& 2.5  & 1  &  1.10 	\\
     		& 5.0  & 0.7&  1.26 	\\

\\
     
\hline  

\\

LBDS 53W091     & 0.01 & 11 &  3.17    \\
		& 0.2  & 10 &  1.26    \\
     		& 0.5  & 8  &  1.32    \\
     		& 1.0  & 3  &  1.21    \\
     		& 1.5  & 3  &  1.30    \\
     		& 2.5  & 1  &  1.37    \\
     		& 5.0  & 0.7&  1.51    \\

\\

\hline  

\end{tabular}
\end{center}

\caption{The results of fitting the near-ultraviolet spectra of the F-stars and the LBDS galaxies using the suite of seven single-metallicity stellar population models. The best fits for each of the single-metallicity models are given. The best fitting models in the case of the F4V star, 53W069 and 53W091 are acceptable fits - very acceptable in the case of the two galaxies. However, the best fitting single metallicity model to the F8V star is not formally acceptable without further relaxation of the assumed errors. See \S3.2 for discussion.}
\end{table}

The fitting process results in a minimum $\chi^2_{\nu}$ corresponding
to an age of 3 Gyr, with a metallicity of 0.2 \Zsolar\ for the F4V
star, HR 4683, and an age of 3 Gyr, with a metallicity of 1.5
\Zsolar for the F8V star, HR 4688. 

Again, the metallicities yielded by our completely independent,
near-UV model-fitting shown in figures 3 and 4 are in very good
agreement with those previously derived from optical indices by
Edvardsson et al. (1993); we have derived metallicities from the
near-UV data which fall within the uncertainties in the
optically-derived values (see figures 3 and 4).

It should be noted that the ages determined by Edvardsson and
co-workers are the present ages of the stars, whereas fitting the
stellar population models to the spectra should yield ages which
correspond to the main-sequence lifetimes of the stars. For this
reason it might be expected that our model-fitting should yield ages
greater than those derived by Edvardsson and co-workers. However, the
models include integration down the main-sequence, as well as post
main-sequence contributions, both of which are likely to lead to an
under-estimate of the true main-sequence turnoff age when fitted to
the spectrum of a single star. Therefore, more important than the
exact values of the derived ages for single stars is the fact that the
age appears to be well constrained despite the availability of seven
different metallicities at each age. In addition, the results of these
stellar population model fits are in good agreement with the results
of \S 3.1, as one would expect if stellar populations are indeed
dominated by MSTO stars at ages $<$ 5 Gyr.

These results therefore demonstrate that by fitting the
single-metallicity models to the near-UV spectra of the F stars, we
have apparently managed to simultaneously determine both age and
metallicity, recovering well-constrained and plausible metallicities
for these two stars.

\begin{figure*}

\centerline{\epsfig{file=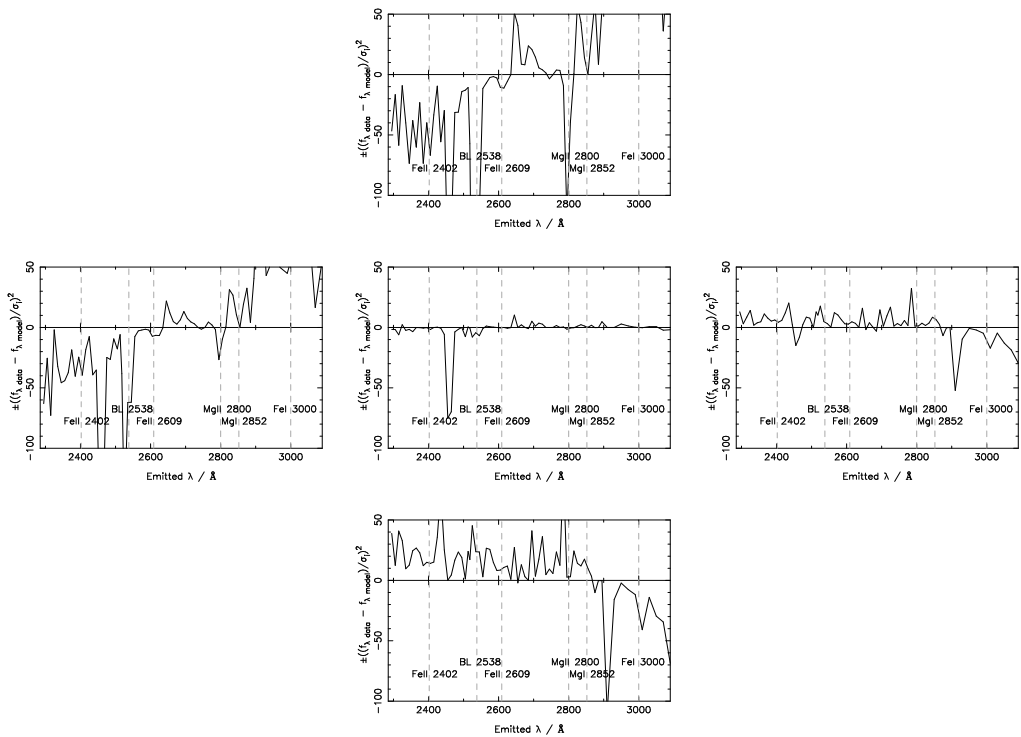,width=11.0cm,angle=-90,clip=}}

\caption{\xs\ residuals for single stellar population fits to the F4V
star, HR 4683. The models fitted have the following ages and
metallicities: top: 3 Gyr, 0.01 \Zsolar; middle: 3 Gyr, 0.2 \Zsolar;
bottom: 3 Gyr, \Zsolar; left: 1 Gyr, 0.2 \Zsolar; right: 7 Gyr, 0.2
\Zsolar. Where $ ( {\rm f}_{\lambda data} < {\rm f}_{\lambda model}
)$, the quantity plotted is $- ( ({\rm f}_{\lambda data} - {\rm
f}_{\lambda model}) / \sigma )^{2}$. The central wavelengths of the
mid-UV spectral features identified by Fanelli et al. (1992), lying
within this wavelength range, are marked on the spectra with dashed
lines. See \S3.2 for discussion.}

\end{figure*}

\begin{figure*}

\centerline{\epsfig{file=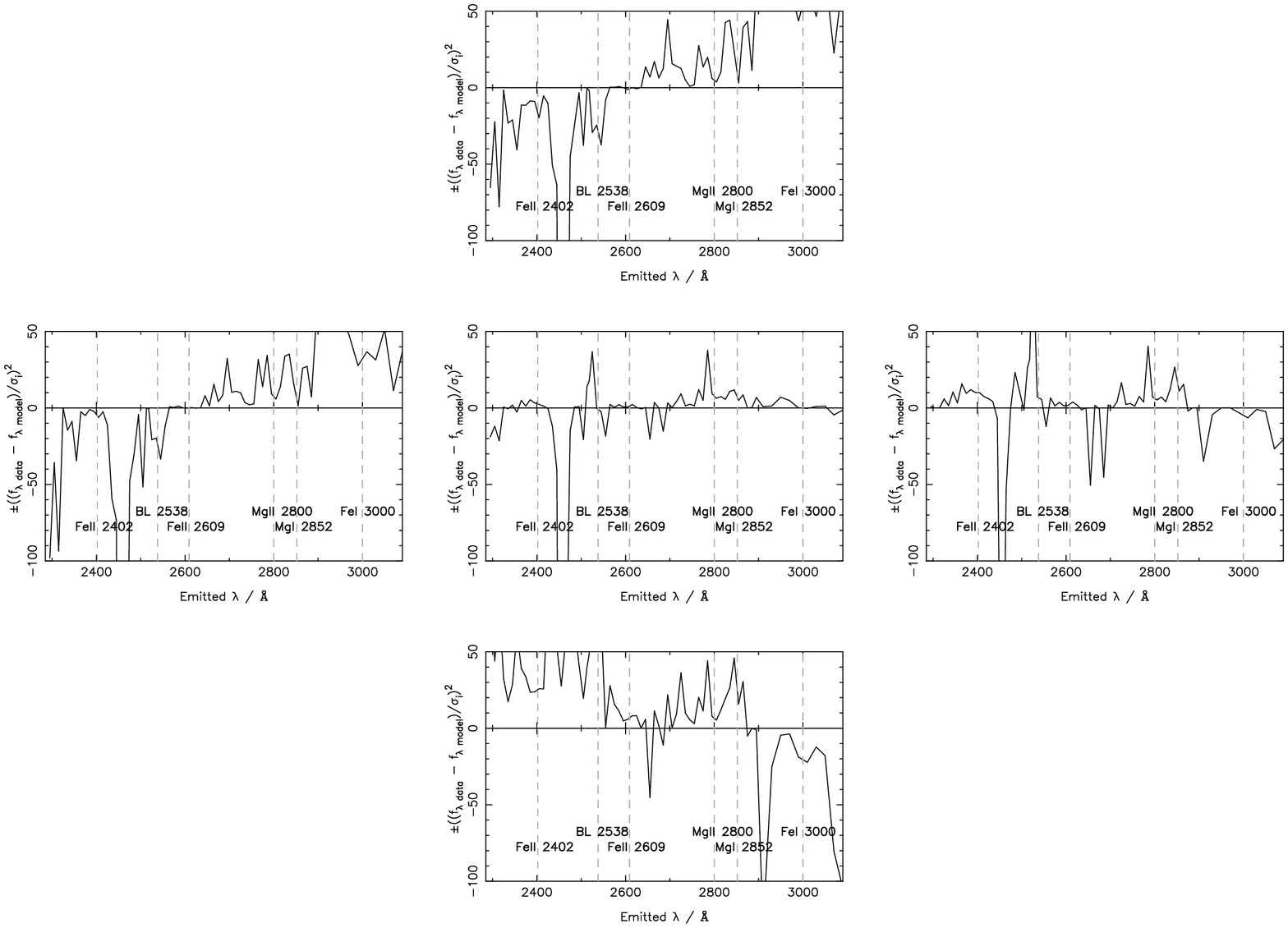,width=11.0cm,angle=-90,clip=}}

\caption{\xs\ residuals for single stellar population fits to the F8V
star, HR 4688. The models fitted have the following ages and
metallicities: top: 3 Gyr, 0.5 \Zsolar; middle: 3 Gyr, 1.5 \Zsolar;
bottom: 3 Gyr, 2.5 \Zsolar; left: 1 Gyr, 1.5 \Zsolar; right: 7 Gyr,
1.5 \Zsolar. Where $ ( {\rm f}_{\lambda data} < {\rm f}_{\lambda
model} )$, the quantity plotted is $- ( ({\rm f}_{\lambda data} - {\rm
f}_{\lambda model}) / \sigma )^{2}$. The central wavelengths of the
mid-UV spectral features identified by Fanelli et al. (1992), lying
within this wavelength range, are marked on the spectra with dashed
lines. See \S3.2 for discussion.}

\end{figure*}

\subsubsection{Age / metallicity determinators}

This success leads naturally to the question of whether any specific
spectral features in this near-UV regime can be identified as
primarily responsible for allow the metallicity and age to be so
effectively determined in this analysis. To explore this we have
derived \xs\ residual plots for a range of single-metallicity stellar
population models fitted to the observed stellar spectra, which are
shown in figures 5 and 6. The central plot is the residual from the best
over-all fit. The metallicity of the model fitted increases downwards,
and the age increases from left to right. Where the (normalised)
observed flux is less than the (normalised) model flux,the quantity
plotted is $- ( {\rm f}_{\lambda model} - {\rm f}_{\lambda data} /
\sigma )^{2}$. The feature at 2450 \ang\ has been masked out in all
fits.

In fact, from these figures it is hard to identify any particular
absorption line as being primarily an indicator of age or of
metallicity. However, there is a clear suggestion that changes in age
and metallicity alter the shape of the residual spectrum in rather
different ways; while changing age alters the overall slope of the
residual continuum level, changing the metallicity introduces stronger
steps into the residual continuum level, especially at $\sim$
2850. This is perhaps not surprising since much of the detailed UV
continuum shape produced by stellar populations around this age is
dominated by extensive line blanketing.

For the sub-solar metallicity star, HR 4683, the spectral features identified by Fanelli et al. (1992), as marked on the plots, are remarkably well reproduced by the best-fitting model spectrum (figure 5, centre plot). The BL 2538 feature is least-well reproduced, although this is better fit by the best-fitting single stellar atmosphere model spectrum. One might expect that, as the spectral features of Fanelli et al. (1992) have been well studied, their flux would be well modelled.

For the super-solar metallicity star, HR 4688, the MgII 2800 and BL
2538 features (offset in the model flux with respect to the central
wavelengths of the features because of the binning resolution) are not
well fitted by the best-fitting single stellar population
model (figure 6, centre plot). It is unlikely that the poor modelling of these features arises
from including post-MS contributions to the model flux, as they are
not well modelled even by the best-fitting single stellar atmosphere
model. The MgII feature at 2800 \ang\ is better fit by the 0.5
\Zsolar\ model. This illustrates the dangers associated with
attempts to determine age and metallicity using only a handful of
absorption features.

\begin{figure*}
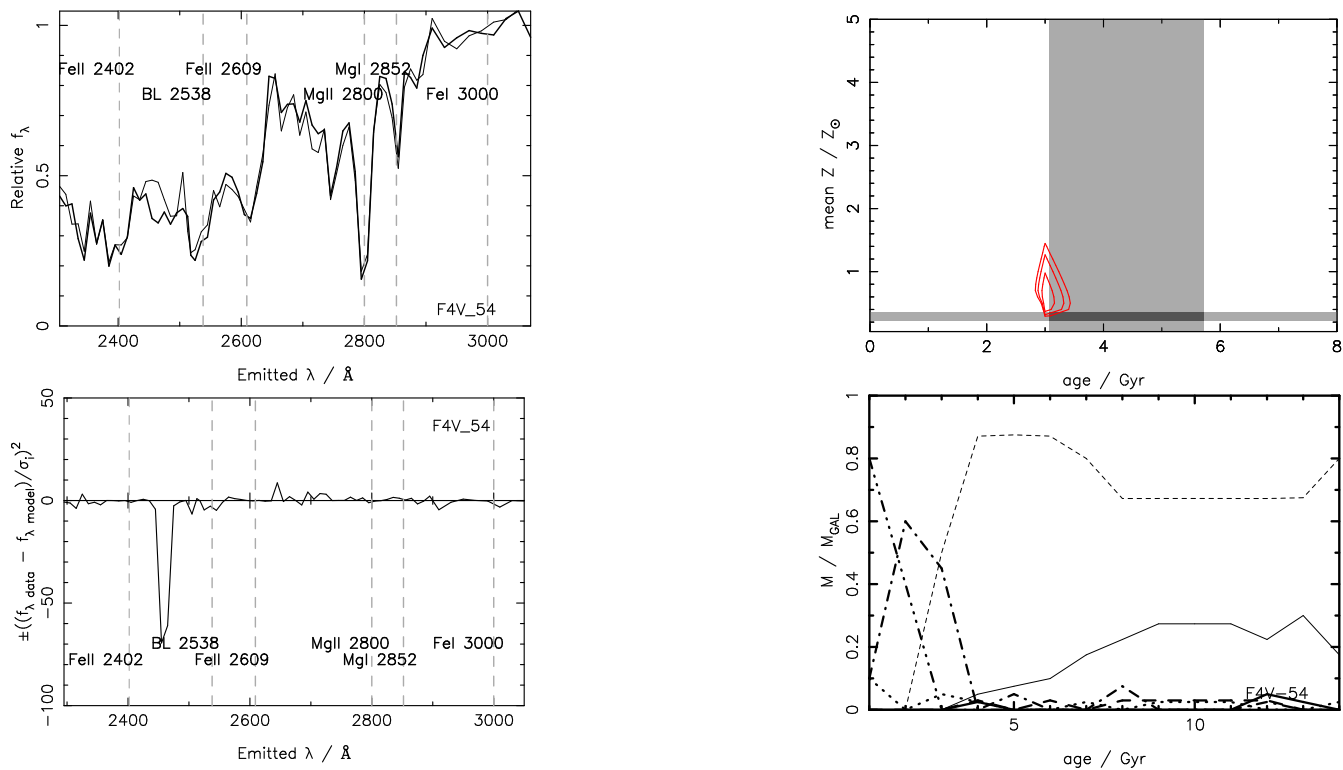


\raggedright{\epsfig{file=F4V_54.epsf,width=5.0cm,angle=-90,clip=}}

\raggedright{\epsfig{file=F4V_54.xspec.epsf,width=5.0cm,angle=-90,clip=}}

\vspace*{-10.0cm}
\raggedleft{\epsfig{file=F4V_54.7zcont.epsf,width=5.0cm,angle=-90,clip=}}

\raggedleft{\epsfig{file=ZevolF4V_54.epsf,width=5.0cm,angle=-90,clip=}}

\caption{ Top left: the best-fitting, seven-component,
mixed-metallicity model (3 Gyr, 0.37 \Zsolar, thin line - see \S3.3
for details) to the spectrum of the F4V star, HR 4683 (thick line).
Bottom left: \xs\ residuals. Where $ ( {\rm f}_{\lambda data} < {\rm
f}_{\lambda model} )$, the quantity plotted is $- ( ({\rm f}_{\lambda
data} - {\rm f}_{\lambda model}) / \sigma )^{2}$. The central
wavelengths of the mid-UV spectral features identified by Fanelli et
al. (1992), lying within this wavelength range, are marked on the
spectra with dashed lines. Top right: contour plots of constant
relative likelihood, for the marginalised distribution of mean
metallicity with age. The contours contain 68.3\%, 90\% and 95.4\%
relative likelihood. The shaded regions in the likelihood plots
represent the estimates of Edvardsson et al., with estimated errors
$\pm$ 30\% in age and $\pm$ 0.10 dex in metallicity.  Bottom right:
fractional contributions (by baryonic mass) to the mixed metallicity
model of the different metallicity components as a function of age ,
i.e. 0.01\Zsolar\ (thin solid), 0.2\Zsolar\ (thin dashed), 0.5\Zsolar\
(dot-dash), \Zsolar\ (dash-dot-dot-dot), 1.5\Zsolar\ (dotted),
2.5\Zsolar\ (solid) and 5.0\Zsolar\ (dashed) and 2.5\Zsolar\
(solid). See \S3.3 for discussion, and table 3 for values of
$\chi^2_{\nu}$.}

\end{figure*}

\begin{figure*}
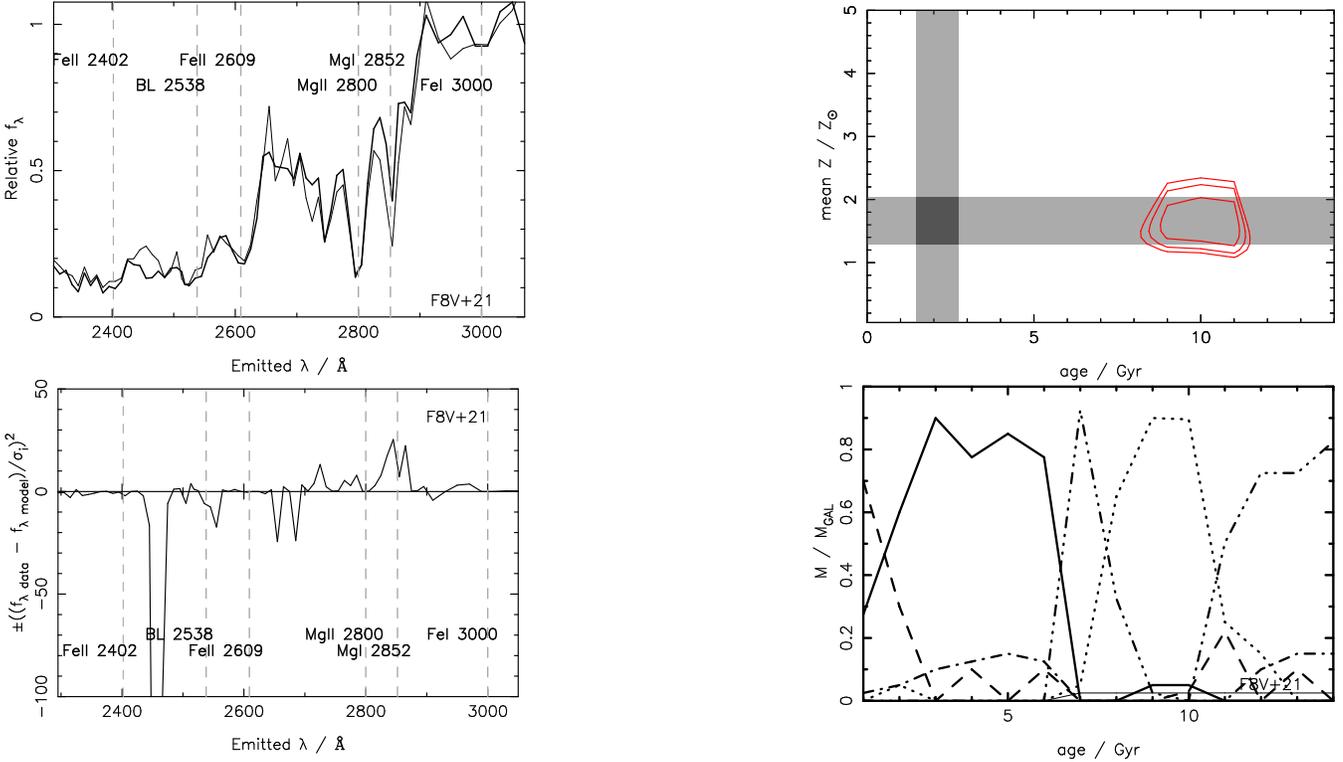

\raggedright{\epsfig{file=F8V+21.epsf,width=5.0cm,angle=-90,clip=}}

\raggedright{\epsfig{file=F8V+21.xspec.epsf,width=5.0cm,angle=-90,clip=}}

\vspace*{-10.0cm}
\raggedleft{\epsfig{file=F8V+21.7zcont.epsf,width=5.0cm,angle=-90,clip=}}

\raggedleft{\epsfig{file=ZevolF8V+21.epsf,width=5.0cm,angle=-90,clip=}}

\caption{ Top left: the seven-component, mixed-metallicity model (9
Gyr, 1.52 \Zsolar, thin line, see \S3.3 for details) fitted to the
spectrum of the F8V star, HR 4688 (thick line), which lies closest to
the parameters corresponding to the minimum interpolated
$\chi^2_{\nu}$. Bottom left: \xs\ residuals. Where $ ( {\rm
f}_{\lambda data} < {\rm f}_{\lambda model} )$, the quantity plotted
is $- ( ({\rm f}_{\lambda data} - {\rm f}_{\lambda model}) / \sigma
)^{2}$. The central wavelengths of the mid-UV spectral features
identified by Fanelli et al. (1992), lying within this wavelength
range, are marked on the spectra with dashed lines. Top right: contour
plots of constant relative likelihood, for the marginalised
distribution of mean metallicity with age. The contours contain
68.3\%, 90\% and 95.4\% relative likelihood. The shaded regions in the
\xs\ plots represent the estimates of Edvardsson et al., with
estimated errors $\pm$ 30\% in age and $\pm$ 0.10 dex in
metallicity. Bottom right: fractional contributions (by baryonic mass)
to the mixed metallicity model of the different metallicity components
as a function of age , i.e. 0.01\Zsolar\ (thin solid), 0.2\Zsolar\
(thin dashed), 0.5\Zsolar\ (dot-dash), \Zsolar\ (dash-dot-dot-dot),
1.5\Zsolar\ (dotted), 2.5\Zsolar\ (solid) and 5.0\Zsolar\
(dashed). See \S3.3 for discussion, and table 3 for values of
$\chi^2_{\nu}$.}

\end{figure*}

\subsection{Mixed-metallicity models}

In figures 7 and 8, we show the results of fitting the
mixed-metallicity model to the HST spectra of the F stars. Each figure
shows the best-fitting model superimposed on the observed near-UV
stellar spectrum, together with the \xs\ residual plots, with the
central wavelengths of the mid-UV spectral features identified by
Fanelli et al. (1992) marked by dashed lines. 

These are followed by
contour plots of constant relative likelihood for the marginalised
distribution of mean metallicity with age. The contours contain
68.3\%, 90\% and 95.4\% relative likelihood. Also included is a fourth
panel showing how the fractional contribution (by baryonic mass) of
the different metallicity components varies as a function of age. The
results are tabulated in table 3.

This time, the interpolated results predict a mean metallicity for the
F4V star of 0.37\Zsolar\, at an age of 3 Gyr, while for the F8V star a
mean metallicity of 1.52\Zsolar, at an age of 9 Gyr, is predicted.

From the \xs\ residual plot in figure, it can be seen that the identified spectral features of HR 4683 are again well fitted by the model. For the super-solar metallicity star, HR 4688, the absorption features BL 2538 and MGII 2800 are much better reproduced by the mixed-metallicity model than by the single-metallicity models.

Fitting this type of mixed-metallicity model arguably provides a
fairer test of our ability to simultaneously determine age and
metallicity. Once again it can be seen that the derived metallicities
and ages are well constrained, and the metallicities are in good
agreement both with the completely independent, optically-based
results of Edvardsson et al., and the results of the previous two
subsections, although, in this case, the predicted age of HR 4688 is
older than might be expected. However, it can be seen in the fourth 
plot of figure 8 that where a single metallicity dominates (at an age
of $\sim$ 3 Gyr) the results are consistent with the results of the
single-metallicity stellar population fits.

\begin{figure}
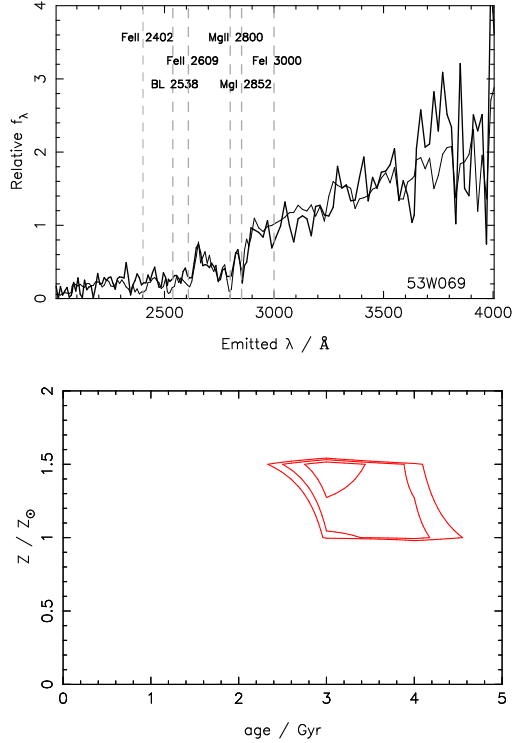

\centerline{\epsfig{file=53W069.epsf.4,width=4.7cm,angle=-90,clip=}}
\vspace*{0.4cm}
\centerline{\epsfig{file=1zcont069.epsf,width=4.7cm,angle=-90,clip=}}

\caption{Top: the best fit single metallicity model (thin line, 1.5
\Zsolar\ model at 3 Gyr, after rejecting ages $>$ 6 Gyr as
cosmologically unreasonable) to the spectrum of the z = 1.43 galaxy,
LBDS 53W069 (thick line). The central wavelengths of the mid-UV
spectral features identified by Fanelli et al. (1992) are marked on
the spectra with dashed lines. Bottom: contours of constant
$\chi^2_{\nu}$ as a function of metallicity and age after
interpolation. The contour levels represent 68\%, 90\% and 95.4\%
confidence. See \S4.1 for discussion, and table 2 for values of
$\chi^2_{\nu}$.}

\end{figure}

\begin{figure}
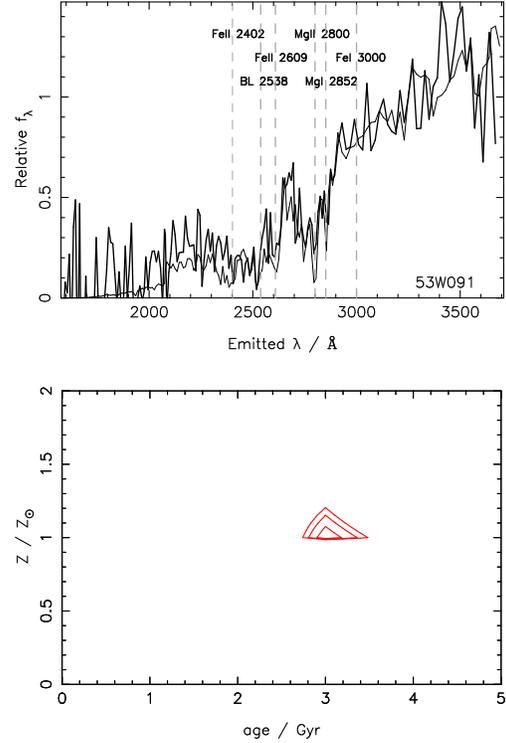

\centerline{\epsfig{file=53W091.epsf.3,width=4.7cm,angle=-90,clip=}}
\vspace*{0.4cm}
\centerline{\epsfig{file=1zcont091.epsf,width=4.7cm,angle=-90,clip=}}

	\caption{Top: the best fit single metallicity model (thin
line, \Zsolar\ model at 3 Gyr) to the spectrum of the z = 1.55 galaxy,
LBDS 53W091 (thick line). The central wavelengths of the mid-UV
spectral features identified by Fanelli et al. (1992) are marked on
the spectra with dashed lines. Bottom: contours of constant
$\chi^2_{\nu}$ as a function of metallicity and age after
interpolation. The contour levels represent 68\%, 90\% and 95.4\%
confidence. See \S4.1 for discussion, and table 2 for values of
$\chi^2_{\nu}$.}

\end{figure}

\section{Red galaxies at z $\simeq$ 1.5}

\subsection{Single-metallicity models}

In figures 9 and 10, we show the results of fitting each of the seven
single-metallicity models in turn to the Keck spectra of the red mJy
radio galaxies LBDS 53W069 (Dunlop 1999, Dey et al. 2002, in
preparation) and LBDS 53W091 (Dunlop et al. 1996; Spinrad et al. 1997)
respectively.  The first plot in each figure shows the best-fitting
model spectrum superimposed on the rest-frame near-UV galaxy
spectrum. The second plot shows contours of constant reduced $\chi^2$
($\chi^2_{\nu}$) as a function of metallicity and age. The contour
levels represent 68\%, 90\% and 95.4\% confidence limits. The results
are presented in table 2. An age of 3 Gyr, with a solar metallicity is
derived for the galaxy LBDS 53W091 (z = 1.55).  For LBDS 53W069 (z =
1.43), ages greater than 6 Gyr are discarded, as they are not possible
at z $\sim 1.5$ in currently-accepted cosmological models. The best
fit age is then 3 Gyr, at a metallicity 1.5 \Zsolar.

In both LBDS 53W069 and 53W091, age and metallicity have been
simultaneously remarkably well constrained (see figures 9 and 10), with
a predicted age for 53W091 between 2.7 and 3.5 Gyr with 95.4\%
confidence, and for 53W069, between 2.3 and 4.6 Gyr at the same level
of confidence. Thus, under the assumption of a single metallicity for
the observed stellar populations in these galaxies, it is clear that
allowing metallicity to vary as a free parameter does not alter the
conclusion of Dunlop (1999), that both these galaxies are $ \sim 3 -
4$ Gyr old, by any more than $\sim$ 1 Gyr, even when, as in the case
of LBDS 53W069, the predicted metallicity is super-solar.

The results confirm the validity of previous arguments in favour of
assuming roughly solar metallicity when modelling LBDS 53W091 (Spinrad
et al. 1997).

\begin{table}

\begin{center}

\begin{tabular}{rrrcr}

\\

\hline

\\

{object}  & { mean Z } & {Best fit age } & {95.4\% age limits } & { \xs$_{\nu}$ 
}       \\
        {  }  & { /{\mbox{\,$\rm Z_{\odot}$}}  } & { / Gyr } & { / Gyr } & {  } 
        \\

\\
     
\hline  

\\

F4V             & 0.37 & 3   & $2.7 - 3.4$  &  1.47  \\
 
F8V             & 1.52 & 9   & $8.0 - 11.5 $ &  1.41  \\

53W069          & 3.78 & $> 6$   & $\geq 4.5$ &  0.78  \\
 
53W091    	& 1.00 & 3   & $2.5 - 4.7$ &  1.25  \\

\\

\hline  

\end{tabular}

\caption{The results of fitting the near-ultraviolet spectra of the
F-stars and the LBDS galaxies with the multi-component,
mixed-metallicity model. Again, the fits are acceptable, although the
fits to the stellar spectra are less good than the fits to the
galaxies, as one might expect from fitting stellar population models
to single stellar spectra.  }

\end{center}

\end{table}

We next investigate the robustness of these results by fitting the
mixed-metallicity model to the galaxy spectra.

\begin{figure}
\centerline{\epsfig{file=53W069.epsf,width=4.7cm,angle=-90,clip=}}
\vspace*{0.4cm}
\centerline{\epsfig{file=53W069.7zcont.epsf,width=4.7cm,angle=-90,clip=}}
\vspace*{0.4cm}
\centerline{\epsfig{file=Zevol53W069.epsf,width=4.7cm,angle=-90,clip=}}

\caption{ Top: the best-fitting, seven-component, mixed-metallicity
model (6 Gyr, 3.78 \Zsolar, thin line) to the spectrum of the
high-redshift (z = 1.43) galaxy LBDS 53W069 (thick line). The central
wavelengths of the mid-UV spectral features identified by Fanelli et
al. (1992) are marked on the spectra with dashed lines. Middle:
contour plots of constant relative likelihood, for the marginalised
distribution of mean metallicity with age. The contours contain
68.3\%, 90\% and 95.4\% relative likelihood. Bottom: fractional
contributions (by baryonic mass) to the mixed-metallicity model of the
different metallicity components as a function of age ,
i.e. 0.01\Zsolar\ (thin solid), 0.2\Zsolar\ (thin dashed), 0.5\Zsolar\
(dot-dash), \Zsolar\ (dash-dot-dot-dot), 1.5\Zsolar\ (dotted),
2.5\Zsolar\ (solid) and 5.0\Zsolar\ (dashed). See \S4.2 for
discussion, and table 3 for values of $\chi^2_{\nu}$.}

\end{figure}

\begin{figure}
\centerline{\epsfig{file=53W091.epsf,width=4.7cm,angle=-90,clip=}}
\vspace*{0.4cm}
\centerline{\epsfig{file=53W091.7zcont.epsf,width=4.7cm,angle=-90,clip=}}
\vspace*{0.4cm}
\centerline{\epsfig{file=Zevol53W091.epsf,width=4.7cm,angle=-90,clip=}}

\caption{ Top: the best-fitting, six-component, mixed-metallicity
model (3 Gyr, \Zsolar, thin line) to the spectrum of the high-redshift
(z = 1.55) galaxy LBDS 53W091 (thick line). The central wavelengths of
the mid-UV spectral features identified by Fanelli et al. (1992) are
marked on the spectra with dashed lines. Middle: contour plots of
constant relative likelihood, for the marginalised distribution of
mean metallicity with age. The contours contain 68.3\%, 90\% and
95.4\% relative likelihood. Bottom: fractional contributions (by
baryonic mass) to the mixed-metallicity model of the different
metallicity components as a function of age, i.e. 0.01\Zsolar\ (thin
solid), 0.2\Zsolar\ (thin dashed), 0.5\Zsolar\ (dot-dash), \Zsolar\
(dash-dot-dot-dot), 1.5\Zsolar\ (dotted), 2.5\Zsolar\ (solid) and
5.0\Zsolar\ (dashed). See \S4.2 for discussion, and table 3 for values
of $\chi^2_{\nu}$.}

\end{figure}

\begin{figure}
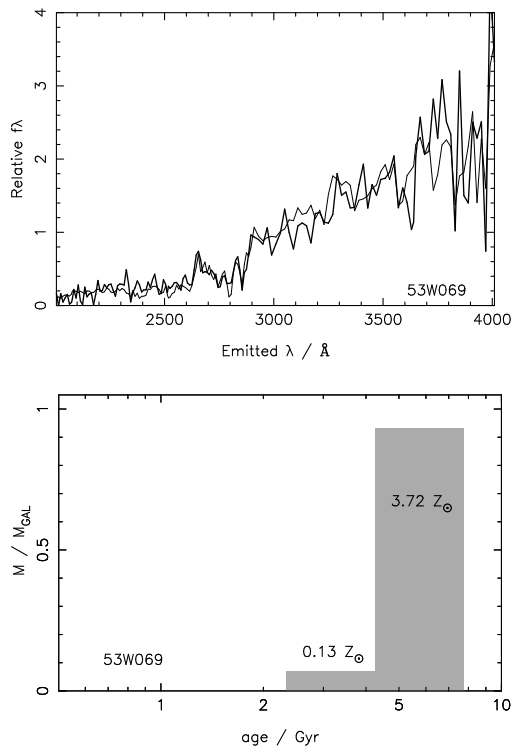

\centerline{\epsfig{file=MOPED069.epsf,width=4.7cm,angle=-90,clip=}}
\vspace*{0.4cm}
\centerline{\epsfig{file=histMOPED069.ps,width=4.7cm,angle=-90,clip=}}

\caption{Top: best-fitting model (thick line) to the spectrum of LBDS 53W069 (thin line) predicted using MOPED (Heavens et al., 2000). Bottom: The predicted stellar population components. The population is dominated by the high metallicity component, with an age $4.25 - 7.72$ Gyr.  }

\end{figure}

\begin{figure}
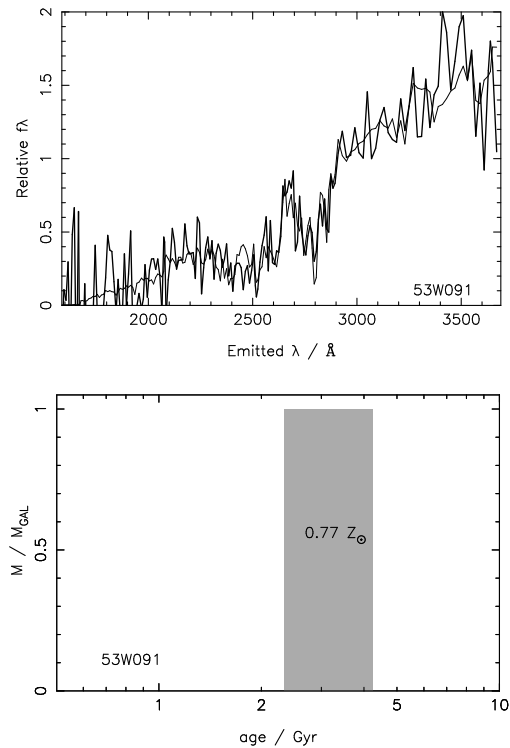

\centerline{\epsfig{file=MOPED091.epsf,width=4.7cm,angle=-90,clip=}}
\vspace*{0.4cm}
\centerline{\epsfig{file=histMOPED091.ps,width=4.7cm,angle=-90,clip=}}

\caption{Top: best-fitting model (thick line) to the spectrum of LBDS 53W091 (thin line) predicted using MOPED (Heavens et al., 2000). Bottom: The predicted stellar population components. The population has negligible contributions from any component except the 0.77\Zsolar\ component, at an age of $2.35 - 4.25$ Gyr.  }

\end{figure}

\subsection{Mixed-metallicity models}

In figures 10 and 11, we show the results of fitting the
mixed-metallicity model to the Keck spectra of the LBDS galaxies.  As
in figures 7 and 8, each figure shows the best-fitting model
superimposed on the rest-frame near-UV spectrum, followed by contour
plots of constant $\chi^2_{\nu}$ as a function of mean metallicity and
age. The contours contain 68.3\%, 90\% and 95.4\% relative
likelihood. Again, a third panel is included showing how the
fractional contribution (by baryonic mass) of the different
metallicity components varies as a function of age. The results are
presented in table 3. Again, ages greater than 6 Gyr are rejected as
impossible in currently accepted cosmological models.

This time the best-fit mean metallicity of 53W069 is 3.78
\Zsolar. Only old ages ($>$ 4.5 Gyr) are allowed, even at high
metallicity. The third plot of figure 11 shows that the
mixed-metallicity model is dominated by contributions from the high
metallicity (2.5 and 5.0 \Zsolar) models. This is a very different
result from the single-metallicity fit, where the age is constrained
at the 95.4 \% level to between $\sim$ 1.9 $-$ 3.2 Gyr, at less than
half the metallicity here.

Despite the relatively high value of best-fitting mass-weighted
metallicity, the interesting thing is that the inferred age of LBDS
53W069 remains old. The second panel in figure 11 demonstrates that,
while the age is not constrained, ages $<$ 4.5 Gyr are prohibited at
the 95.4\% confidence level.

For 53W091, the mean metallicity derived from this mixed-metallicity
modelling is reasonably well constrained, and is in fact precisely
solar.  Not surprisingly, therefore, the best-fit age remains $\simeq$
3 Gyr (c.f. table 2), with ages $<$ 2.5 Gyr formally rejected at the 95.4\%
confidence level. Again, ages $>$ 6 Gyr have been excluded on cosmological grounds.

Is this lower age limit on 53W091 robust, or are better fits possible
at lower ages with even higher metallicities than those investigated?
For this galaxy, the bottom plot in figure 12 shows that the highest
metallicity component is only dominant at very young ages, where the
statistical quality of the fit is completely unacceptable. A
sufficient region of parameter space has been explored for the age
limit $>$ 2.5 Gyr to be regarded as robust. Again, it should be noted
that where a single metallicity component dominates the fit, the
results are consistent with the single-metallicity model fits.

\subsection{MOPED results}

In all of the above, we have assumed that the galaxies have formed the
bulk of their stellar populations, of whatever metallicity, in a
single coeval starburst. To check the extent to which this
simplification is appropriate, it is clearly desirable to explore a
larger parameter space, in which star formation history is allowed to
vary freely. We have therefore refitted the galaxy spectra using the
MOPED algorithm (Heavens et al., 2000). In this approach, star
formation is divided into several bins in time; the bins have equal
width in log space, and allow widths small enough to account for the
recent formation of stars in giant molecular clouds ($10^7$
years). The height of each bin, allowed to float as a free parameter,
represents the mass turned into stars at the corresponding epoch. The
metallicity of each bin is allowed to take any value from 0.01 to 5
\Zsolar. In this way, we have the most model-independent prescription
for describing a stellar population which is a composite of simple
stellar populations with arbitrary metallicity, age and star formation
rate. Since it is obvious that we cannot explore a parameter space
with 1000 variables, we constrain ourselves to 12 bins in star
formation history. With the help of MOPED (Heavens, Jimenez \& Lahav,
2000, Reichardt, Jimenez \& Heavens, 2001), we are able to explore
this huge (25 dimensional) parameter space very efficiently, in only a
few seconds.

Figures 13 and 14 show the best fitting value found searching the
hyper-likelihood surface\footnote{The hyper-likelihood surface is so
large in this case that not all of it can be explored at once. Instead
we chose 2000 random starting points to find the global minimum, which
is shown in figures 13 and 14, for 53W069 and 53W091
respectively.}. The results for 53W069 yield a population dominated by
an old ($4.25 - 7.72$ Gyr), high-metallicity (3.72 \Zsolar) component,
together with a small (7.0\% by baryonic mass), younger component with a
metallicity of 0.13 \Zsolar. For 53W091, we can see an old component
($2.35 - 4.25$ Gyr) with a metallicity of 0.77 \Zsolar\ and negligible
contributions from other populations, in good agreement with our
previous findings. This, therefore, confirms that for both 53W091 and
53W069, the ages of the dominant populations remain old and apparently
coeval, even when the star formation history is allowed to float
freely. Unsurprisingly, therefore, the quality of the best fits
achieved by MOPED are no better than those achieved with the single
burst models explored in the earlier section of this paper.

It should be noted that the MOPED algorithm includes a parameter which
allows for reddening of the spectra by dust. For both 53W091 and
53W069, the MOPED fits find a negligible contribution from dust. As
the MOPED results are in good agreement with the results from the
mixed-metallicity model spectral fits, we are justified in ignoring
dust in the spectral fits.

\section{Conclusion}
The main conclusions of this work can be summarized as follows.

First, we have demonstrated that, when fitted to high-quality, near-UV
spectra of F stars, the synthetic near-UV spectra produced by the
variable-metallicity models of Jimenez et al. (1998) are sufficiently
accurate to extract well-constrained estimates of metallicity in
accord with the values determined from the study of well-established
optical indices. This suggests that it should, in principle, be
possible to lift the age-metallicity degeneracy when fitting such
models to high-quality optical spectra of high-redshift galaxies.

Second, fitting these variability-metallicity models to the Keck
optical (rest-frame near-UV) spectrum of the red, $z \simeq 1.5$
galaxy 53W091 we find that it is dominated by a 3-Gyr stellar
population of solar metallicity, in good agreement with
the original conclusions of Dunlop et al. (1996).

Third, the fit to the even redder z $ \simeq 1.5$ galaxy 53W069 has
been able to constrain the metallicity well, but the age has not been
convincingly determined. Given the success of the fits to the other
spectra, where the old, 5 \Zsolar\ model does not make any significant
contribution, this is most likely to represent some uncertainty in the
very highest metallicity model at old ages. However, better quality UV
data and / or the addition of rest-frame optical data may serve to
resolve this problem.

The main finding of importance for cosmology is that, even after
this exploration of the effects of varying metallicity, the best-bet age
of the oldest known galaxies at $z \simeq 1.5$ remains 3 Gyr, with ages
younger than 2 Gyr now more strongly excluded than before.

For $H_0 = 70 {\rm km s^{-1} Mpc^{-1}}$ an Einstein-de Sitter universe
is younger than 3 Gyr at $z \simeq 1.5$, and even the hard lower limit
for coeval star formation of 2 Gyr translates into star-formation
activity prior to $z = 8$.  For a flat universe with $\Omega_m = 0.3$
and $\Omega_{\Lambda} = 0.7$, the best-bet age of 3 Gyr translates
into $z = 5$, while the hard lower limit of $>2$ Gyr translates into
$z > 3$. The added flexibility in the star formation history
investigated using the MOPED algorithm allows some later star
formation, but the bulk of the stars ($>$ 90\%) must still be formed
at high redshift.

Our results therefore continue to argue strongly against an
Einstein-de Sitter universe, and favour a $\Lambda$-dominated universe
in which star formation in at least LBDS 53W091 is completed somewhere
in the range $z = 3 - 5$. Such a conclusion is in accord with recent
results from sub-mm surveys which suggest that star-formation activity
in radio galaxies, and potentially all massive ellipticals is largely
completed at $z > 3$ (Archibald et al.  2001; Dunlop
2001).

\vspace*{1.0cm}

\noindent 
{\bf ACKNOWLEDGEMENTS}\\
   \\
We would like to thank Sally Heap for making the F star data available to us. Louisa Nolan acknowledges the support of a PPARC Studentship. James
Dunlop acknowledges the enhanced research time afforded by the award of a 
PPARC Senior Fellowship.
Raul Jimenez acknowledges the support of a PPARC Advanced Fellowship.
   \\
   \\
   \\
   \\
   \\
{\bf REFERENCES}\\
Archibald, E., Dunlop, J.S., Hughes, D.H., Rawlings, S., Eales, S.A., Ivison, R.J., 2001, MNRAS, 323, 417\\
Baker N., \& Temensvary S., 1966, in: Tables of Convective Stellar Models \\
Bruzual G., Magris G.C., 1997, The Ultraviolet Universe at Low
and High Redshift : Probing the Progress of Galaxy Evolution, College 
Park, MD, edited by William H. Waller et al., New York, American Institute of Physics. Also AIP Conference Proceedings, v.408., p.291\\
Chambers K.C., Charlot S., 1990, ApJ, 348, L1\\
Charlot S., Worthey G., Bressan A., 1996, ApJ, 457, 626\\
Dunlop J.S., 1999, in 'The Most Distant Radio Galaxies', Proceedings of the colloquium, Amsterdam, 15-17 October 1997, Royal Netherlands Academy of Arts and Sciences. Edited by H. J. A. R$\o$ttgering, P. N. Best, and M. D. Lehnert., p. 71\\
Dunlop, J.S., 2001, 'Sub-mm clues to elliptical galaxy formation',In: Deep Millimetre Surveys, eds. Lowenthal, J. \& Hughes, D.H., World Scientific, p. 11\\ 
Dunlop J.S., Guideroni B., Rocca-Volmerange B., Peacock J.A., Longair M.S., 1989, MNRAS, 240, 257\\
Dunlop J.S., Peacock J.A., Spinrad H., Dey A., Jimenez R., Stern D., Windhorst R, 1996, Nature, 381, 581\\ 
Edvardsson B., Andersen J., Gustafsson B., Lambert D.L., Nissen P.E., Tomkin J., 1993, A. \&\ A. 275, 101\\
Fanelli M.N., O'Connell R.W., Burstein D., Wu C-C., 1992, ApJS, 82, 197\\
Gustafsson B., Bell R.A., Eriksson K., Nordlund A., 1975, A\&A 42, 407\\
Heap S.R. et al., 1998a, ApJ, 492, L131\\
Heap S.R., Lanz T.M., Brown T., Hubeny I., 1998b, in 'The Hy-Redshift Universe: Galaxy Formation and Evolution at High Redshift', Proceedings of a conference held in Berkeley, CA, 21-24 June, 1999. ASP Conference Proceedings, Vol. 193, Edited by Andrew J. Bunker and Wil J. M. van Breugel, p. 167 \\
Heavens, A.F., Jimenez, R., Lahav, O., 2000, MNRAS, 317, 965\\
Helling C., J$\o$rgensen U.G., Plez B., Johnson H.R., 1996, A\&A, 315, 194\\
Iglesis C.A., \& Rogers F.J., 1996, ApJ, 464, 943\\
Kotoneva E., Flynn C. \& Jimenez R., 2002, MNRAS, in press (astro-ph/0203118)\\
Jimenez R., J$\o$rgensen U.G., Thejll P., MacDonald J., 1995, MNRAS, 275, 1245\\
Jimenez R., \& McDonald J., 1996, MNRAS, 283, 721 \\
Jimenez R., Thejll P., J$\o$rgensen U.G., MacDonald J., Pagel B., 1996, MNRAS, 283, 721\\
Jimenez R., Padoan P., Juvela M., Bowen D.V., Dunlop J.S., Matteucci F., 2000, ApJ, 532, 152\\
Jimenez R., Padoan P., Matteucci F., Heavens A.F., 1998, MNRAS, 229, 123\\
J$\o$rgensen, U.G., 1991, A\&A, 246, 118\\
Kurucz R., 1992, in: The Stellar Populations in Galaxies, ed. B Barbuy, A Renzini, 225\\
Lilly S.J., 1998, ApJ, 333, 161\\
Magris G.C., Bruzual G., 1993, ApJ, 417, 102\\
Mihalas D., 1978, in: Stellar Atmospheres, San Francisco: W.H. Freeman, 1978\\
Nolan L.A., Dunlop J.S., Jimenez R., 2001, MNRAS, 323, 385\\
Reichardt C., Jimenez R. \& Heavens A.F., 2001, 327, 849\\
Reimers D., 1975, Mem. Soc. Roy. Sci. Li$\acute{e}$ge, 8,369\\
Salpeter E.E, 1955, ApJ, 121, 161\\
Spinrad H., Dey A., Stern D., Dunlop J.S., Peacock J.A., Jimenez R., Windhorst R, 1997, ApJ, 484, 581\\
VandenBerg D.A., 1985, ApJS 58, 711\\
Worthey G., 1994, ApJS, 95, 107\\
Yi S., Brown T.M., Heap S., Hubeny I., Landsman W., Lanz T., Sweigart A., 2000, ApJ, 533, 670\\

\end{document}